\documentclass[letterpaper]{sig-alternate}

\usepackage[ruled]{algorithm2e}
\usepackage{stmaryrd}
\usepackage{graphicx}
\usepackage{color}
\usepackage{comment}
\usepackage{url}
\usepackage{wrapfig}

\newcommand{\sysname}{SINk}

\newcommand{\superscript}[1]{\ensuremath{^{\textrm{#1}}}}

\def\wu{\superscript{*}}
\def\wg{\superscript{\dag}}

\begin{document}
% --- Author Metadata here ---
\conferenceinfo{ARM 2015, } {December 07-11, 2015, Vancouver, BC, Canada} 
 \CopyrightYear{2015}
 \crdata{978-1-4503-3733-5/15/12} 
 \clubpenalty=10000 
 \widowpenalty = 10000
% --- End of Author Metadata ---

\title{\sysname : A Middleware for Synchronization of Heterogeneous Software Interfaces}

\numberofauthors{1} %  in this sample file, there are a *total*
\author{
% 1st. author
\alignauthor Mohammad Hosseini\wu, Yu Jiang\wu, Poliang Wu\wu, Richard B. Berlin Jr.\wu\wg, Lui Sha\wu\\
       \affaddr{~}\\
%       \affaddr{\wg Carle Foundation Hospital\wg}\\
       %  \sharedaffiliation
  \begin{tabular}{ccc}  
    \wu Department of Computer Science & & \wg Department of Surgery \\
    University of Illinois at Urbana-Champaign (UIUC)    & & Carle Foundation Hospital \\
  \end{tabular}
  ~\\
      \email{\{shossen2, jy1989, wu87, rberlin, rberlin, lrs\}@illinois.edu}
 %2nd. author
%\alignauthor Richard B. Berlin Jr.\wu\wg\\
%      \affaddr{Carle Foundation Hospital \\ University of Illinois at Urbana-Champaign}\\
%      \email{rberlin@illinois.edu}
}

\date{26 November 2014}
\maketitle

\begin{abstract}
Software is everywhere. The increasing requirement of supporting a wide variety of domains has rapidly increased the complexity of software systems, making them hard to maintain and the training process harder for end-users, which in turn ultimately demanded the development of user-friendly application software with simple interfaces that makes them easy, especially for non-professional personnel, to employ, and interact with.

However, due to the lack of source code access for third-party software and the lack of non-graphical interfaces such as web-services or RMI (Remote Method Invocation) access to application functionality, synchronization between heterogeneous closed-box software interfaces and a user-friendly version of those interfaces has become a major challenge. In this paper, we design \sysname \footnote{A demo illustrating how our middleware works in practice is available at http://publish.illinois.edu/mdpnp-architecture/?p=639}, a middleware that enables synchronization of multiple heterogeneous software applications, using only graphical interface, without the need for source code access or access to the entire platform's control. \sysname\ helps with synchronization of closed-box industry software, where in fact the only possible way of communication is through software interfaces. It leverages efficient client sever architecture, socket based protocol, adaptation to resolution changes, and parameter mapping mechanisms to transfer control events to ensure the real-time requirements of synchronization among multiple interfaces are met. Our proof-of-concept evaluation shows there is in fact potential usage of our middleware in a wide variety of domains. 

\end{abstract}



%\keywords{Mobile Video Streaming, Energy Awareness, DASH}

\section{Introduction}
``Software is eating the world!''~\cite{andreessen2011software}. Our dependency on software is continuously increasing, %Most of today's personal uses and industry services are realized, at least in part, by means of %software systems. 
and it is said that 60-90\% of production in the automotive domain for example, is done by software
systems \cite{complexSoftware}. Many products and industrial services that would have traditionally been realized through ``hardware'', are now realized purely via ``software solutions''. 
Overall, one way or another, human-in-the-loop software systems in various domains are getting more and more complex, as they operate within a complex ecosystem of libraries, models, protocols and devices, and require human interaction \cite{complexSoftware}. 
%However, software systems require changes over time in response to new applications, technologies, or paradigms, as a result of applying patches to found security, logical, or performance-related vulnerabilities, or because of changes in resource availability and reconfiguration of the underlying execution platform. However, how to effectively and timely adapt to ecosystem changes while guaranteeing applications can consistently keep on working securely and effectively in spite of these changes has become a key challenge. Unfortunately, the lack of automated mechanisms to restructure and transform applications when changes do occur leads to high software maintenance costs.
The interfaces of many platform-dependent software, such as industrial controller simulators (e.g. Mitsubishi PLC x7 \cite{park2006development}) and healthcare systems (e.g. Laerdal's SimMan Patient Simulator \cite{simman}) for example, are sometimes hard to manage and lack user-friendliness. Therefore, third-parties are pushed to develop simple-to-use, and more user-friendly and maneuverable interfaces for those applications, which in fact motivates the need for co-simulation among different interfaces. While the graphical user interfaces are easy to develop, there has been a significant demand on interface-to-interface synchronization of heterogeneous software interfaces. 

\begin{figure}[!tp]
\centering
%\fbox{
\includegraphics[width=1\columnwidth]{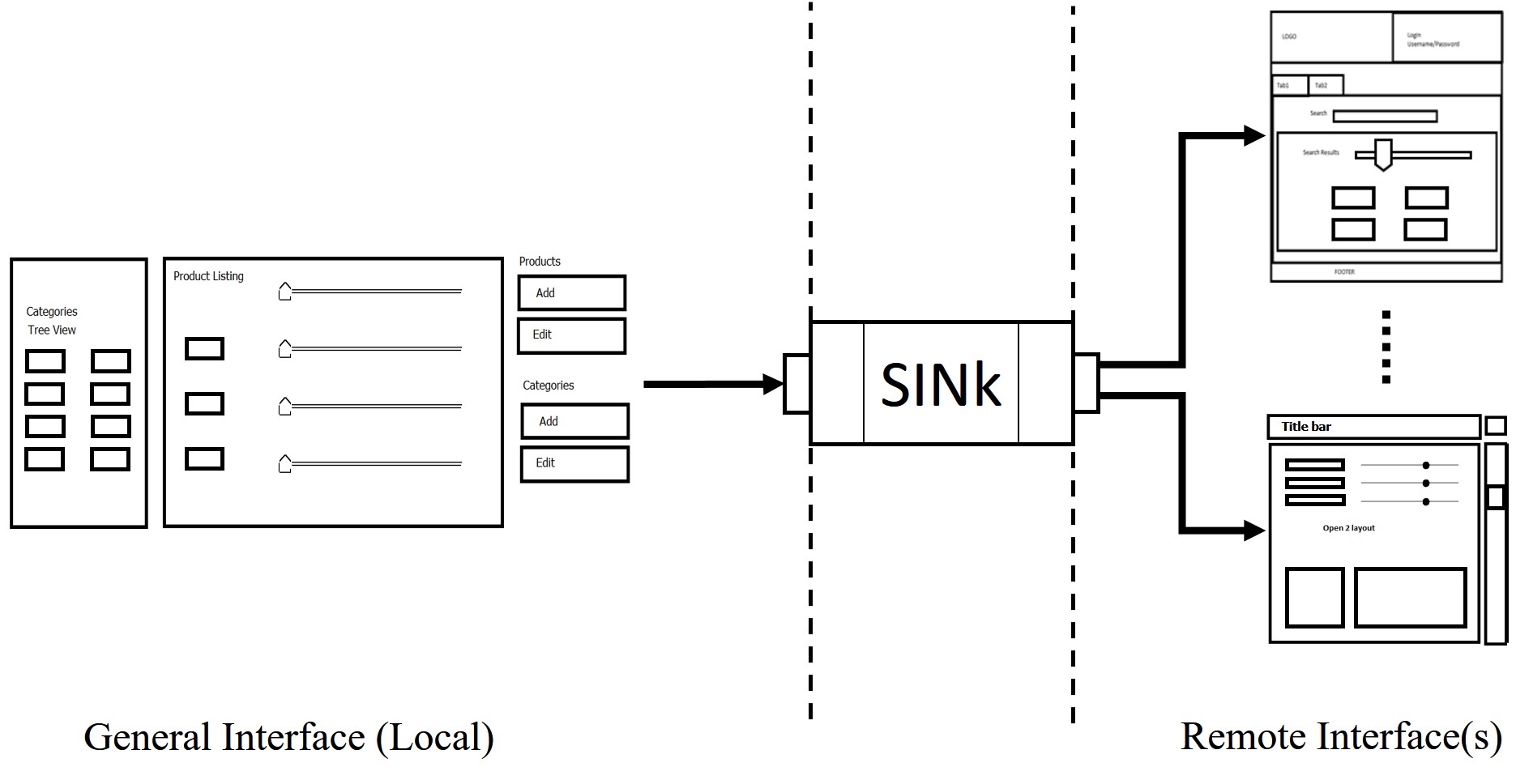}
%}
\caption{The \sysname\ workflow. Multiple graphical interfaces (right-side) are remotely synchronized with a single interface (left).}
\label{Fig:Workflow}
\end{figure}

Unfortunately, existing tools such as those spanning from remote desktop applications and desktop sharing to collaborative software applications lack support of interface synchronization, and only provide access to applications simply through desktop screen sharing and manual control by users. 

In this paper, we describe \sysname, a middleware that enables interface-to-interface synchronization and automatic remote control of heterogeneous graphical interfaces. The design of this middleware is mainly motivated by connecting and synchronizing heterogeneous applications with homogeneous functionality, but different graphical interfaces, over a network or the Internet, as presented in Figure \ref{Fig:Workflow}. Once the interfaces are connected through the established client-server connection, our middleware allows users to automatically synchronize multiple applications, remotely, without physical access or visual view of the remote desktops, as if they were sitting right in front of the remote applications. In this manner, we enable the user to control any remote applications and automatically perform actions such as opening and closing windows and tabs, pushing buttons, applying keystrokes, and updating strings, values, and checkboxes, using only their graphical interfaces and without causing mismatches between them. \sysname ~is adaptive in the sense that it can accommodate varying platform features such as changes in display screen resolutions, as it can dynamically adapt itself to locate pixel values \textit{relative} to any resolution. Moreover, \sysname 's automated mechanisms achieved through \textit{interface-only} control incurs a high degree of flexibility, and can effectively adapt to ecosystem changes when reconfiguration of application rapidly occurs. That leads to significant reduction in heterogeneous software maintenance costs.

Technically speaking, the \sysname ~middleware leverages efficient architecture, protocol, and parameter mapping mechanisms to transfer control events, while at the same time ensuring consistency, bandwidth saving, platform independence and the fulfillment of real-time requirements for synchronization. In summary, \sysname %\vspace{-0.1cm}
\begin{itemize}
\item automatically performs remote control as opposed to manual control by users,\vspace{-0.2cm}
\item does not require visual view of remote desktop, thus providing significant bandwidth savings, \vspace{-0.2cm}
\item does not require source code or non-graphical interfaces (such as web services or RMI) access to remote applications,\vspace{-0.2cm}
\item performs synchronization in real-time,\vspace{-0.2cm}
\item is platform independent.
\end{itemize}
To the best of our knowledge, no single middleware currently exists that achieves synchronization among heterogeneous applications in a coherent way. Moreover, \sysname ~can further assist software engineers to build a single user-friendly interface as a general application interface for front-end interaction, or to realize co-simulation among multiple heterogeneous graphical interfaces. 
%Figure \ref{Fig:Workflow} illustrates how \sysname\ works in practice.

\section{Related Work}
\sysname\ is conceptually similar to the notion of \textit{mediators} underlying emergent connectors \cite{mediator1, mediator2, mediator3} such as Enterprise Service Bus \cite{esb} as the concept of a ``connectivity middleware" is common between the two. However, \sysname\ is fundamentally different as the design goal of mediators is to enable the composition of pervasive networked systems, protocol mediation, and interoperability in distributed systems as opposed to remote interface-based synchronization. The most related tools to \sysname\ are remote desktop and desktop sharing software, which allow a personal computer's desktop environment to be run remotely on one system, while being displayed on a separate client device. Microsoft's Remote Desktop Connection \cite{remoteDesktopConnection}, Apple Remote Desktop \cite{apple}, and Chrome Remote Desktop \cite{chrome} for example, allow users to remotely connect to a computer from another computer, therefore providing access to programs and files by visually controlling the keyboard and mouse and relaying the graphical screen over a network. Similarly, desktop sharing applications and collaborative software such as Microsoft's Lync \cite{lync} and TeamViewer \cite{teamviewer} provide desktop access to a remote machine running the same software, helping users to remotely control and share a desktop, with the additional option of video conferencing services. However, not only are these applications manual and user-controlled, but the remote desktop software and desktop sharing applications also act in a \textit{computer-to-computer} manner and have computer-wide access. This is not easily applied to the synchronization among platform dependent applications, such as flight control and autopilot systems in drones~\cite{autopilot}, automatic remotely-controlled construction machinery in smart-grids~\cite{fukushima}, and co-simulation of heterogeneous production and ERP software in the automotive industry~\cite{automotive}.

In \sysname\ , on the contrary, the notion of access is platform independent and lightweight because it is \textit{application-to-application} or \textit{interface-to-interface}. Furthermore, control and input parameters are directed \textit{automatically} into the remote graphical application interfaces residing on the remote computers, thus synchronizing multiple interfaces and allowing users to need only control a \textit{single} interface. In addition, \sysname\ eliminates the unnecessary need to visually share the desktop views, hence allowing for significant bandwidth savings by avoiding the real-time encoding and transfer of desktop views, especially crucial for power-limited mobile devices \cite{mmsys12,mmsys13}. Moreover, remote users have no ability to modify the shared content and resources whatsoever, and are only passively controlling remote interfaces.

\section{Design of the Middleware}
\sysname\ is implemented through a mapping system as well as a communication system accomplished through a client/server architecture. The client is installed on the local computer running the local application and then connects to the server component, which is installed on the remote computer. During \sysname\ sessions, all corresponding keystrokes and mouse clicks are registered as if the users were actually sitting in front of the remote computers and performing tasks on the remote interfaces. We implemented \sysname\ in Java that can be deployed on any platform running Java Virtual Machine (JVM), including Linux and Windows. Therefore, JVM is a base requirement, making the compiled code platform-independent. We have designed a list of APIs for the users, such as performing a remote connection, specifying control attributes, and transferring parameter values.

\subsection{Middleware Structure}

\begin{figure*}[!tp]
    \centering
        \includegraphics[trim=0cm 0cm 5cm 2.5cm, width=.7\textwidth]{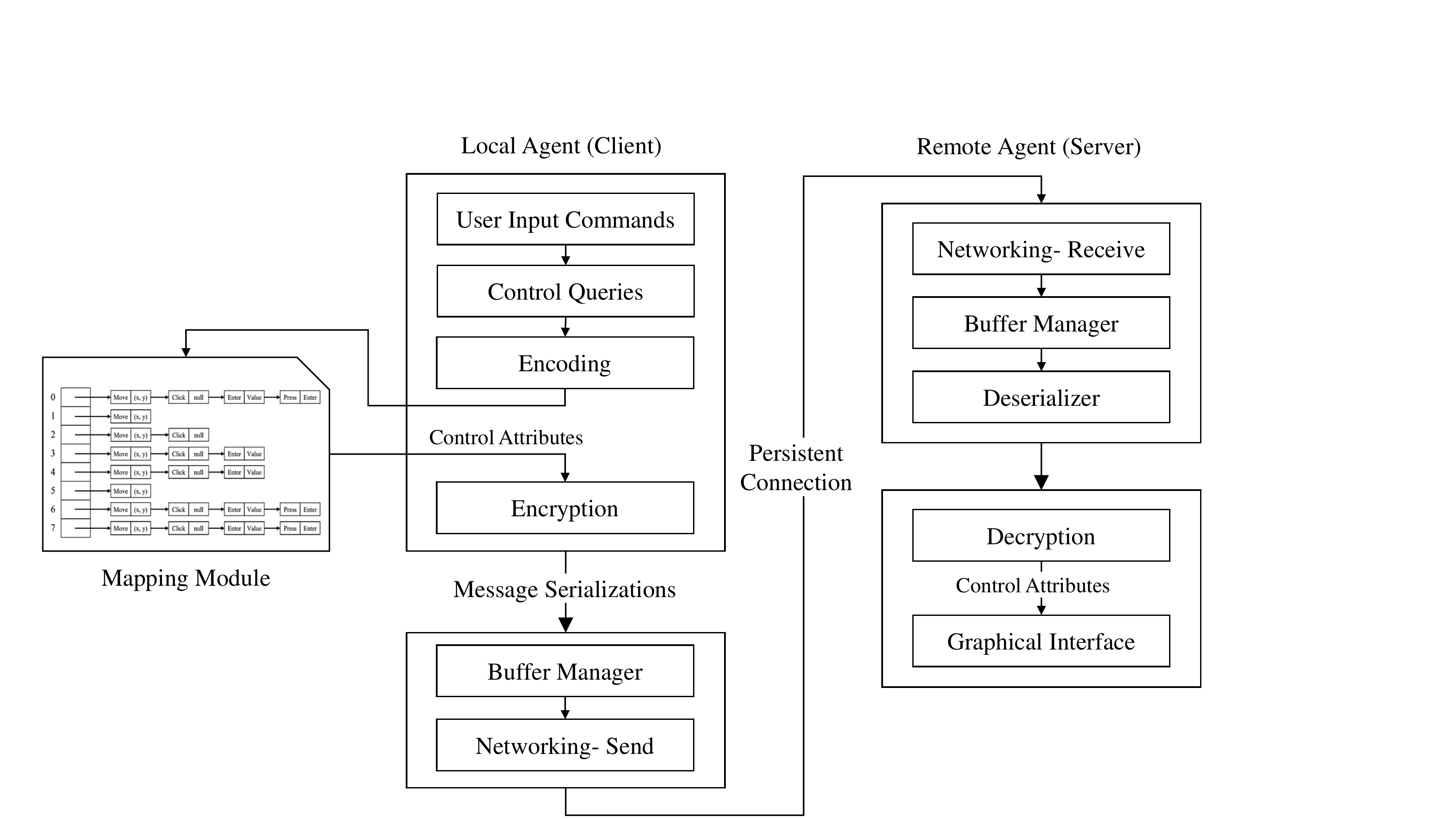}
    \caption{The overall structure of \sysname .}
    \label{Fig:Structure}
\end{figure*}

\begin{figure}[!t]
\centering
\vspace{.5cm}
\includegraphics[width=\columnwidth]{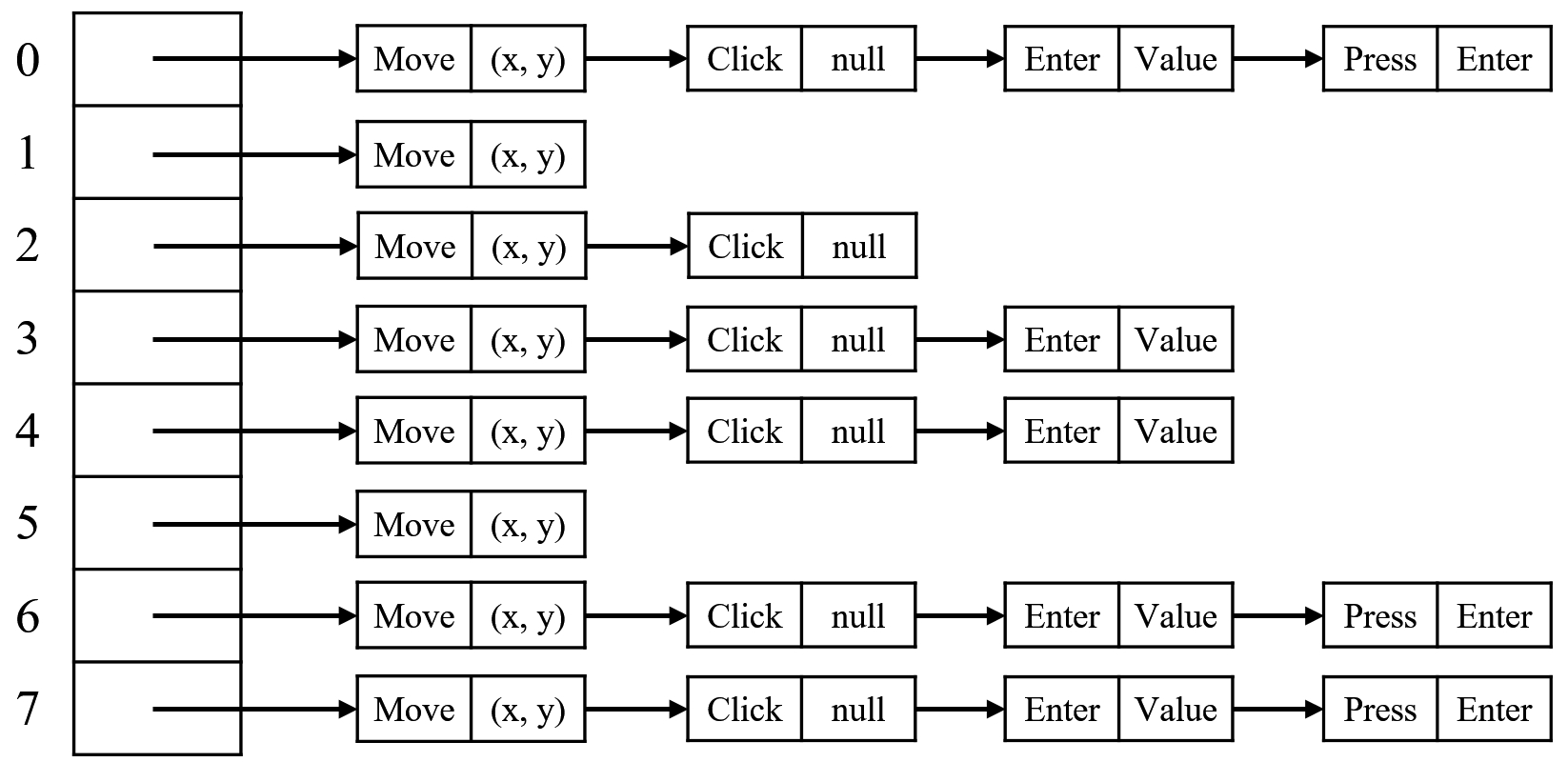} %
\caption{An example mapping module.}
\label{Fig:map}
\end{figure}

\sysname\ consists of three major components: a local agent (control client) residing on the local machine, a mapping module, and a remote agent (control server) residing on each remote machine, as illustrated in Figure \ref{Fig:Structure}. The local agent communicates with the remote agents through a secure persistent message exchange communication system. Users' control inputs are received by the local agent, encoded to a specific message format, and are then directed to the mapping module. 

The mapping module is pre-configured with interface control attributes to provide a particular set of interface functions on each of the remote interfaces. This happens by performing transformation of local control inputs on the local platform to remote interface control attributes corresponding to the remote graphical interfaces on the remote platform, thereby allowing remote synchronization with each remote interface. This is similar to a remote desktop connection, but it happens automatically and with no need for sharing the desktop view or visual control with users. The output data originating from the mapping module consists of control attributes, which are then encrypted with the AES 128-bit symmetric cipher in electronic codebook (ECB) mode, buffered, serialized, and then transferred to the remote agent via the persistent socket connection. While placement of the mapping module as a centralized module on the local machine is more convenient for updates, auditing, and security reasons, it is not yet a hard requirement. The module can be placed separately on each individual remote machine alternatively.

The control messages are deserialized and decrypted once received at the remote agent. Remote synchronization between the local interface and remote interfaces is performed via interface control functions in accordance with the control attributes received through the communication channel.  Although currently synchronization is only one-way (from local to the remote interfaces), without loss of generality, \sysname ~can be reconfigured so that changes and results on remote interfaces be synchronized back and displayed on the local interface for any possible adaptation purposes.

\subsection{Customized Client-Sever Architecture }
From an engineering point of view, unlike a regular client-server connection such as those in chat systems with the client looping to read the responses, our middleware tool must also support sporadic message transfer but with no connection termination. However, it also needs to maintain a live and permanent connection after each transfer in order to incur minimum latency.

To address the requirement above, we customized a low-overhead persistent client-server connection over TCP/IP throughout the running session rather than setting up a new connection for each transfer. This maintains the stability of the socket connections by initially creating a connection at the beginning of each session, and occasionally sending a message given the system's input. To enable that, we wrapped the client socket connection around a thread, and use a blocking queue to wait for messages. A single sender queue exists throughout the application, therefore using a singleton pattern. On the other side, performing a $\texttt{read()}$ function causes the thread to block forever. To address that, we use a special type of thread that calls a specific method repetitively at specified periods and read time-out that can be used to post a message, a ping message, every so often, which improves the stability of connections while also relaxing problems associated with closing the applications due to calling the $\texttt{close()}$ function.

\subsection{Data Structures and Rules for Mapping}

The mapping module works on the principle of key-value store and hashing, composed of a combination of hash-map and 2-dimensional linked-list data structures, which is used to simulate user interaction and control the graphical interfaces pre-configured with mouse and keyboard events. To store key-value pairs, we used the first dimension of the 2D linked-list as a bucket to store key objects corresponding to encoded user inputs, while the second dimension is used to store values, corresponding to an ordered list of interface control attributes such as necessary mouse clicks and key presses that must be executed on the remote interfaces to perform identical actions. Similar to a regular \textit{HashMap}, the mapping module's $\texttt{get(Key k)}$ method calls \textit{hashCode} method on the key inputs, and applies returned \textit{hashValue} to its own static hash function to find a bucket location where keys and values are stored. Figure \ref{Fig:map} shows an overview of an example mapping module. For example, the first entry of the map as shown in Figure \ref{Fig:map}, executes the following chain of events:

\begin{enumerate}
\item \textit{Move} the mouse pointer to a specific 2D coordinate on the display screen (given x and y coordinates as the horizontal and vertical addresses of any pixel, respectively),
\item Perform a mouse \textit{click} event on current pointer (we implemented click event as a combination of mouse left button's \textit{press} and \textit{release} events, with an intermediate delay of 200 ms),
\item \textit{Enter} a specific value or number in the current position. This requires it to iteratively \textit{press} multiple specific keys on the keyboard, and
\item \textit{Press} the ``Enter" key on the keyboard.
\end{enumerate}

\begin{wrapfigure}{R}{0.2\textwidth}
\setlength{\columnsep}{2pt}%
%    \vspace{-.2cm}
  \centering
  \hspace{-.3cm}
    \includegraphics[scale=1]{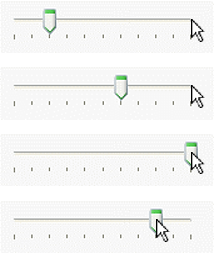}
    \caption{Interface control process for an example horizontal slider bar.}
    \label{fig:slider}
\end{wrapfigure}
Our implementation of the mapping module imposes a one-time overhead, and it can be reusable. Therefore, if the application's user interface changes considerably, only the mapping module is updated, incurring minimum cost so the automation does not need to be rewritten.

\begin{figure*}[!htbp]
\centering

\includegraphics[width=.65\textwidth]{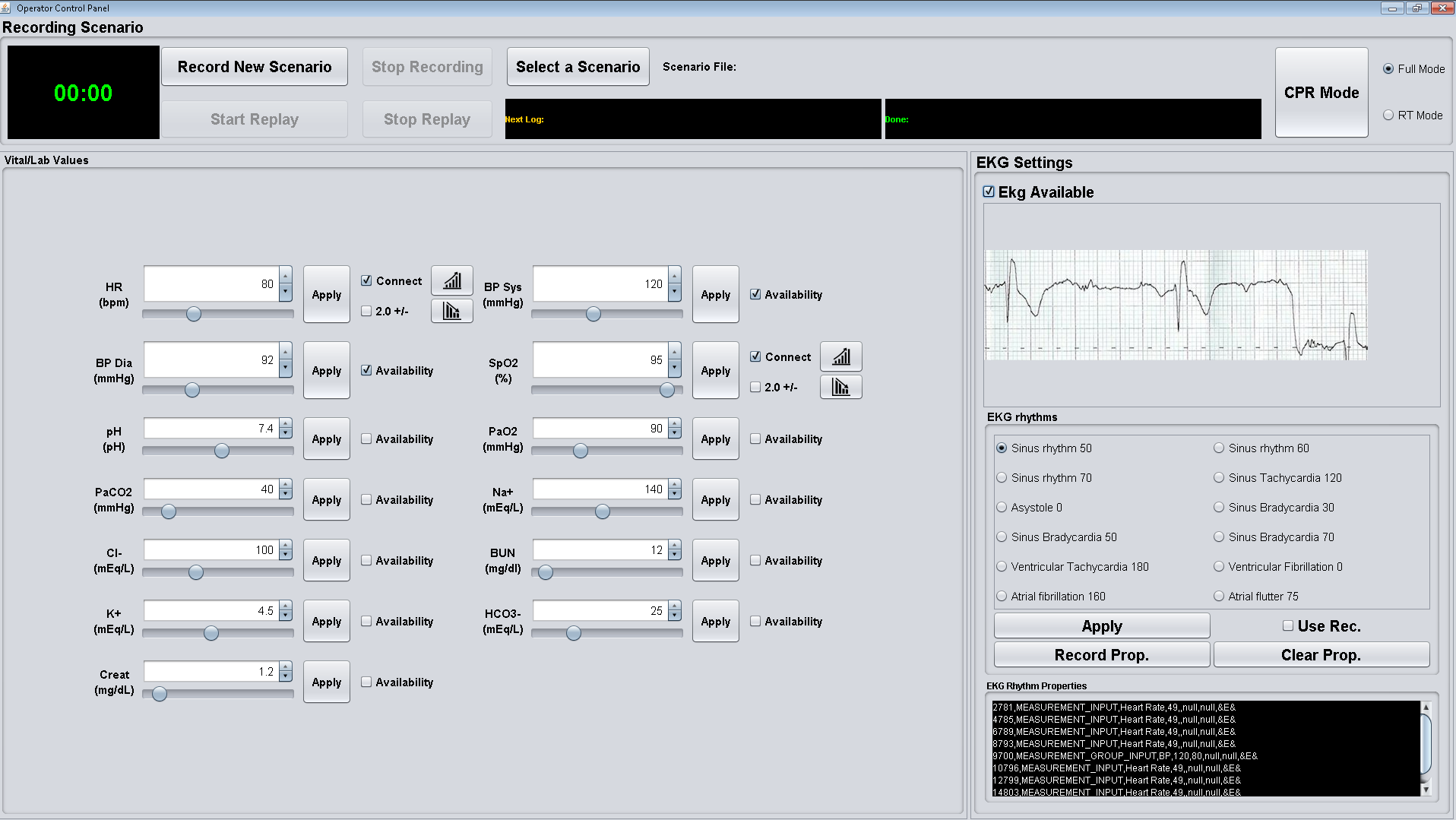}
~\\~\\
\includegraphics[width=.65\textwidth]{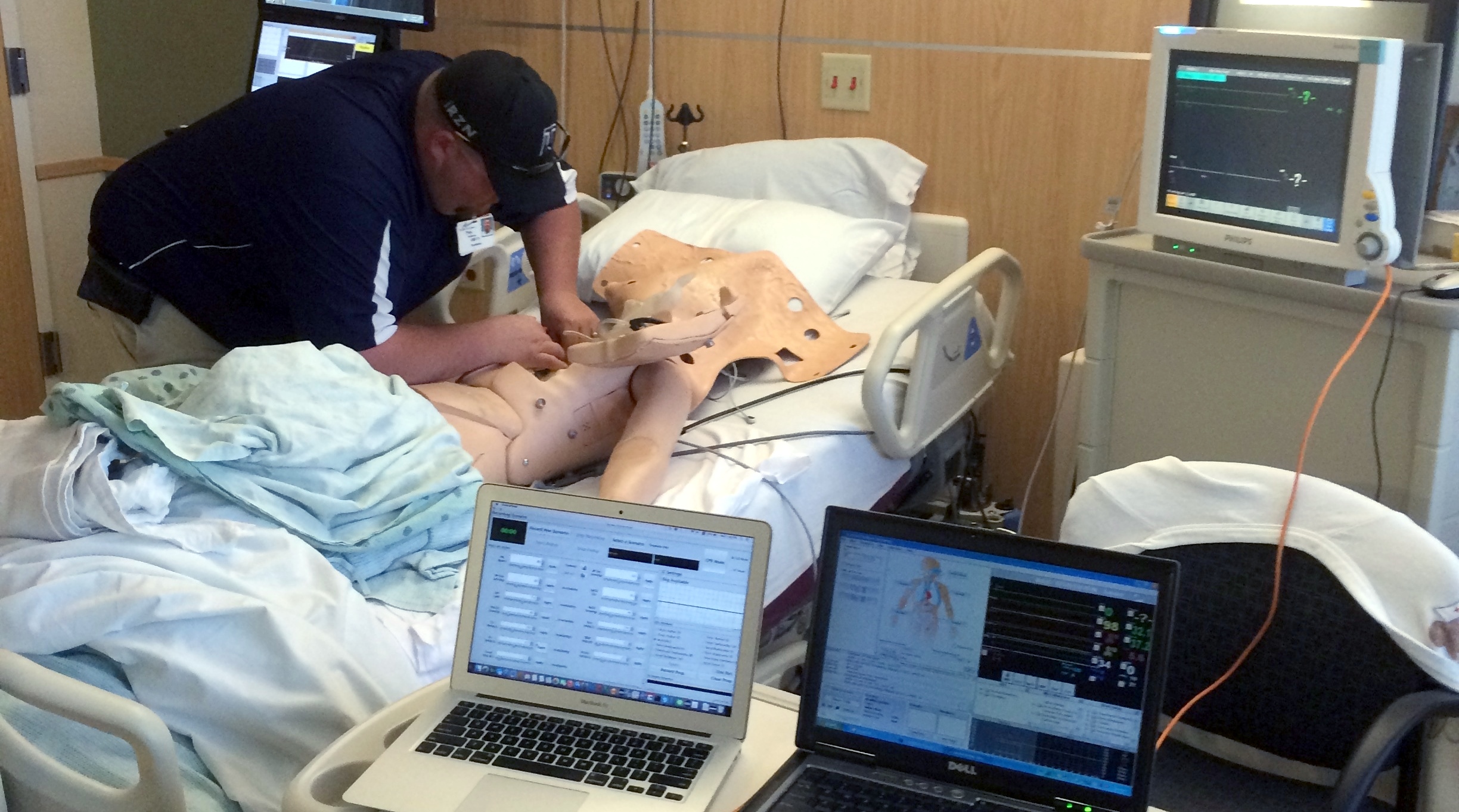}
~\\~\\
\includegraphics[width=.65\textwidth]{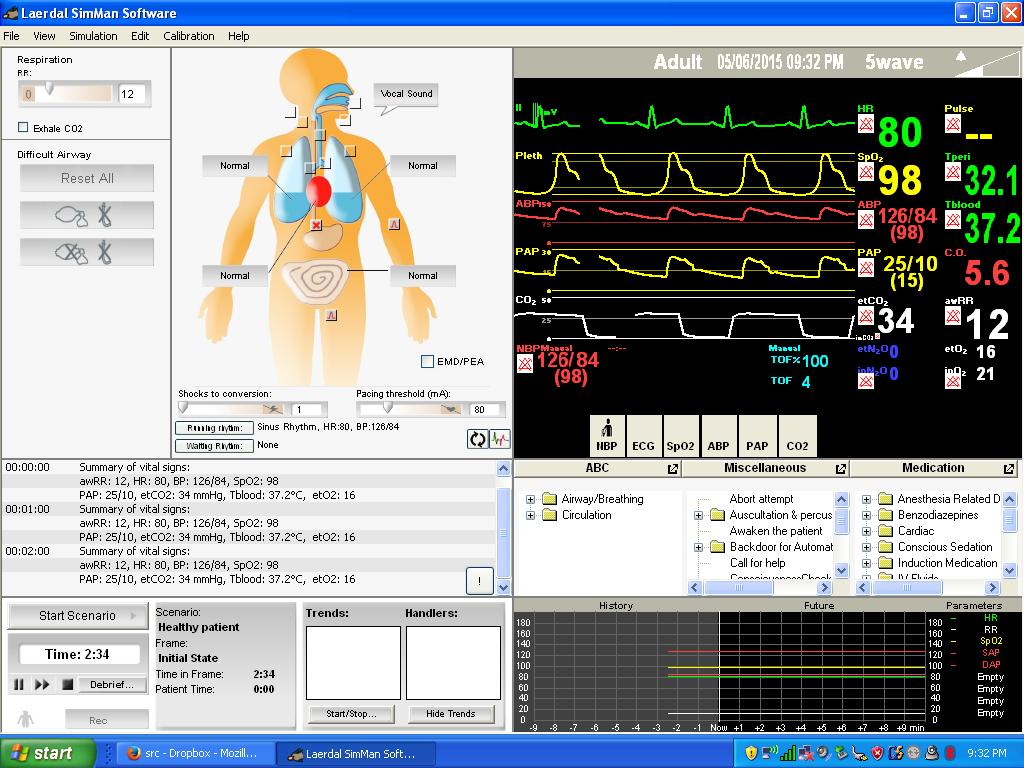}

\caption{Real platform testing (Middle). Local interface (Top). Remote closed-box application (Bottom).}
\label{Fig:Experiment}
\end{figure*}

The graphical interface control attributes are implemented as a series of sequential mouse and keyboard events. While many of the graphical user interface components are elementary actions and are straightforward to control through events originating from multiple registered mice and keyboards, such as moving the mouse pointer to a specific coordinate location, clicking, pressing a key or entering a value, interesting challenges exist when controlling or adjusting some interface components such as scroll bars and slider bars. Let's take a horizontal slider bar for example. To control a slider bar to set a new value, use of a mouse drag event is infeasible as the initial position of the slider knob is unknown for the mouse pointer to hover on. While the current position of the slider knob is not known, the coordinates of minimum or maximum value endpoints are known on the horizontal bar. The design trick to address this challenging issue is to first move the mouse pointer to the coordinate corresponding to either endpoint, and then perform mouse clicks multiple times to push the slider pointer on the track towards the specific endpoint (the maximum number of mouse clicks is deterministic- in our experiments the number was four). Once the mouse pointer is on the slider knob, a mouse drag event is then performed to move the knob to the desired pre-determined position corresponding to the desired control value. Figure \ref{fig:slider} (top to bottom) visually illustrates the control process. Although graphical component functions and views are subject to operating system, design language and layout variants, mouse and keyboard events can be registered for our control mapping purposes without loss of generality.

\section{Evaluation}

%\begin{figure*}[!htbp]
%\centering
%\fbox{
%\includegraphics[height=0.13\textheight]{Experiment-Figure/our1.png}~~\includegraphics[height=0.18\textheight]{Experiment-Figure/Experiment.jpg}~~\includegraphics[height=0.13\textheight]{Experiment-Figure/untitled0.jpg}}
%\caption{Real platform simulation and testing. (Left) local interface. (Right) remote closed-box application.}
%\label{Fig:Experiment}
%\end{figure*}

We have evaluated and tested \sysname\ rigorously over our industrial case study conducted in collaboration with Carle Foundation Hospital~\cite{carle}, on a real platform where 138 synchronization requirements were specified to synchronize two medical simulator software products. The requirements were inspected multiple times with developers, researchers, and physicians to ensure that specific functional requirements are satisfied.

The evaluation platform is presented in Figure \ref{Fig:Experiment}, with the closed-box simulator software as the remote interface on the bottom, and the local interface illustrated on the top. The closed-box software is SimMan's~\cite{simman} advanced patient simulator shipped with a laptop running Windows XP, which controls a SimMan medical manikin used for basic and advanced life Support skill assessment. The SimMan's simulator software allows observation, recognition, and modification of most vital signs which are used in emergency medicine, fed directly to the manikin itself as well as a patient monitor. The local interface likewise, is a patient simulator locally developed for nurses and physicians as a part of a best practice medical system to perform the most relevant medical interventions according to the medical guidelines and protocols. The local patient simulator features a simple, user-friendly, and easily-operated graphical interface with straightforward and uncomplicated control functions to help nurses and physicians avoid complications of using the SimMan's patient simulator. As an example, the local interface incurs a \textit{single} step including 10 parameters to modify the running heart rhythm, whereas the SimMan's patient simulator involves 9 steps, requiring the user to audit 57 different parameters. With \sysname, the input values are only controlled through a single user-friendly interface, and are automatically synchronized with those corresponding to the SimMan's simulator, thereby relieving the users from confusing complications and removing the need to double-enter the input values on a second interface. All 138 synchronization requirements are accomplished correctly.

Apart from the case study and the important benefits resulting from using \sysname, our middleware was specially regarded for its automation role. Prior to applying our middleware, a technician was hired to replicate, and manually perform the control functions on the SimMan's simulator as a way to synchronize the user-friendly patient simulator with the SimMan's simulator. Thenceforth, automatic synchronization was achieved, removing the human from the loop. Overall, we have received positive feedback from the experts using the middleware. The qualitative feedback we have received is promising, suggesting the middleware might be applicable to large sets of requirements and extended to domains such as co-simulation of heterogeneous production and ERP software in the automotive industry~\cite{automotive}.
\newpage
\section{Conclusion and Future Work}
In this paper, we presented \sysname , an adaptive middleware tool that performs interface synchronization automatically, remotely, and without physical access or visual view of remote interfaces, as if users were sitting right in front of the remote software. We tested and evaluated \sysname\ on a real platform, and showed that apart from daily personal applications, there are in fact many potential uses of our middleware in industry services that can not be realized by other means.

We are currently working on an interface attribute recorder that can capture and log interface control inputs on local interfaces and be fed directly into the mapping module, to strengthen the automation and the scalability of the middleware. We can also exploit image segmentation and energy-efficient texture recognition techniques to learn type and position of graphical components on software interfaces, especially when aimed at interfaces on power-constrained mobile devices \cite{shervin1, shervin2, mm12}. In the future, we also plan to systematically evaluate \sysname\ using quantitative metrics.
%\newpage
\section{Acknowledgments}
This research is funded in part by NSF CNS 13-29886 and in part by Navy N00014-12-1-0046.
%\newpage
\bibliographystyle{abbrv}
\bibliography{sigproc}

\end{document}